\documentclass[conference]{IEEEtran}
\usepackage{etex}

\usepackage{stfloats}

\usepackage{pstricks}
\usepackage{lipsum}
\usepackage[numbers,sort&compress]{natbib}

\usepackage{amssymb}
\usepackage{pgfplots}
\usepackage{graphicx}
\graphicspath{{C:/Users/laureano/Documents/ICC_2015-MIMO_k-mu_shadowed/Figuras/}}
\pgfplotsset{compat=1.3}

\ifCLASSINFOpdf

\else

\fi
\usepackage[cmex10]{amsmath}

\begin{document}
%
\title{On Some Unifications Arising from\\ the MIMO Rician Shadowed Model}


\author{\IEEEauthorblockN{Laureano Moreno-Pozas and Eduardo Martos-Naya}
\IEEEauthorblockA{Dpto. Ingenier\'ia de Comunicaciones, ETSI Telecomunicaci\'on\\
Universidad de M\'alaga, M\'alaga, 29071,  Spain\\ Email: \{lmp, eduardo\}@ic.uma.es}
}


\maketitle

\begin{abstract}
This paper shows that the proposed Rician shadowed model for multi-antenna communications allows for the unification of a wide set of models, both for multiple-input multiple output (MIMO) and single-input single output (SISO) communications. The MIMO Rayleigh and MIMO Rician can be deduced from the MIMO Rician shadowed, and so their SISO counterparts. Other SISO models, besides the Rician shadowed proposed by Abdi et. al., are included in the model, such as the $\kappa$-$\mu$ defined by Yacoub, and its recent generalization, the \mbox{$\kappa$-$\mu$} shadowed model. Moreover, the SISO \mbox{$\eta$-$\mu$} and \mbox{Nakagami-$q$} models can be seen as particular cases of the MIMO Rician shadowed. The literature already presents the probability density function (pdf) of the Rician shadowed Gram channel matrix in terms of the well-known gamma-Wishart distribution. We here derive its moment generating function in a tractable form. Closed-form expressions for the cumulative distribution function and the pdf of the maximum eigenvalue are also carried out.
\end{abstract}


%
\IEEEpeerreviewmaketitle

\section{Introduction}

The Rician fading model \cite{Rician} was proposed in order to characterize scenarios where there is a dominant signal, oftentimes called line-of sight (LOS) signal, whose power is much stronger than the power of the rest of the signals received due to reflections, widely referred to as scattering waves.

With the aim of also including the large-scale propagation effects, a Rician shadowed model was then presented by Loo, where the LOS signal suffers from a perturbation in the amplitude which follows a log-normal distribution \cite{Loo}.

Although the model presented by Loo is validated with channel measurements, the statistical characterization of such model is not simple since the probability density function (pdf) is given in an integral form. Instead, a new Rician shadowed model was proposed in \cite{Abdi}, where the shadowing amplitude follows a Nakagami-$m$ distribution. In fact, this new fading model can be easily characterized without compromising the accuracy when fitting real channel measurements \cite{Abdi}.

Recently, the Rician shadowed proposed in \cite{Abdi} was generalized under the name of $\kappa$-$\mu$ shadowed fading model \cite{Paris} or shadowed $\kappa$-$\mu$ \cite{CottonShadowed}, and exhibits excellent agreement when compared to measured underwater acoustic \cite{Paris, ParisConf} and body centric communications fading channels \cite{CottonShadowed, Cotton}.

In the literature, the statistical characterization of the aforementioned channel models is usually tackled on a single-link fashion, i.e., for a single-input single-output (SISO) communication system.

However, since modern communication systems like \mbox{Wi-Fi} standards or 4G always use several antennas at both the transmitter and receiver sides, i.e. multiple-input multiple-output (MIMO) systems, the statistical characterization of MIMO systems is of extreme interest. 

The study of the impact of MIMO diversity in the capacity of flat-fading Rayleigh channels was studied for a long while \cite{WintersCo, WintersCapacity, Telatar} and was finally analytically evaluated for spatially correlated Rayleigh channels in \cite{ChianiRayleigh}. The MIMO Rician was also deeply studied in the literature \cite{Khalighi, MckayBounds, AlouiniCapacityRician}, showing that its performance analysis is much more complicated than the MIMO Rayleigh case. Moreover, when trying to extend other types of models in order to be employed in MIMO communications, we tackle some analytic problems. For instance, the MIMO \mbox{Nakagami-$q$} model is still an open problem in the literature, being only partially characterized in \cite{MIMO2x2, Kumar}, leading to comlicated and not very tractable expressions. In fact, random matrix models for fading channels other than Rayleigh or Rician are scarce in the literature \cite{MIMO2x2, Kumar, Alfano, Mckay}.

In this paper, we revisit the random matrix model for a Rician fading model with Gamma-variate average matrix presented in \cite{Alfano}, which we here directly call MIMO Rician shadowed for the first time. We show that the versatility of the model has not been exploited yet to the full extent possible, since this model allows for a wide unification: i) for MIMO communications, it unifies the MIMO Rayleigh and MIMO Rician models, and ii) for SISO communications, it unifies the one-sided Gaussian, Rayleigh, Nakagami-$m$, Nakagami-$q$ and Rician, together with their general counterparts, i.e, the $\kappa$-$\mu$ and $\eta$-$\mu$ \cite{Yacoub}, the Rician shadowed \citep{Abdi} and the $\kappa$-$\mu$ shadowed \cite{Paris, CottonShadowed}. Moreover, like its univariate counterpart, this model can be also used when its parameters takes non-integer values.

Although the pdf of the Gram channel matrix can be found in \cite{Alfano}, we here present for the first time the moment generating function (mgf) of such model in closed-form. Moreover, we derive the cumulative density function (cdf) and pdf of the maximum eigenvalue distribution for a shadowing power matrix with equals eigenvalues, which are the respective distributions of the outage probability and the maximum output SNR pdf of a maximum ratio combining system \cite{Tse2000, Dighe2003}.

This paper is structured as follows. In Section II, we introduce some preliminary results needed in our following derivations. In Section III, the MIMO Rician shadowed model is again introduced and we derive the mgf of the Gram channel matrix. In Section IV, we prove the wide unification that model allows. In Section V, the cdf and pdf of the maximum eigenvalue distribution of the random matrix model are derived for an shadowing with equals eigenvalues. In Section VI, we present some numerical results. Finally, conclusions are drawn.

Throughout this paper, matrices are denoted in bold uppercase. The matrix $\mathbf{I}_p$ symbolizes the $p\times p$ identity matrix, while $\mathbf{0}_p$ is the $p\times p$ null matrix. When the operator $\vert \cdot \vert$ is used around a matrix, it indicates the determinant of that square matrix; otherwise, it is the complex modulus. The matrix $\mathbf{\bar{A}}$ is the expectation matrix of $\bf{A}$. The conditional matrix $\bf{A\vert B}$ means the matrix $\bf{A}$ given matrix $\bf{B}$. The operator tr$(\cdot)$ represents the matrix trace while etr$(\cdot)$ is the exponential of the matrix trace. The super-index $\cal{^H}$ means the conjugate transpose and the symbol $\sim$ expresses \emph{statistically distributed as}. If $\bf{H}$ is a $p\times n$ matrix, we refer to $\bf{H}^{\cal{H}}\mathbf{H}$ as its Gram matrix. Finally, $\mathbf{\mathcal{V}(\cdot)}$ is the Vandermonde determinant and $\mathbf{A}>0$ indicates that matrix $\mathbf{A}$ is positive definite.

\section{Preliminaries}

\emph{Definition 1: Noncentral Complex Wishart Matrix.}\\
The $n\times n$ matrix $\mathbf{W}$ is a noncentral complex Wishart matrix with $p$ degrees of freedom ($p\geq n$), covariance matrix $\bf{\Sigma}$ and noncentrality matrix $\mathbf{\Theta}$, i.e.,  $\mathbf{W}\sim{\mathcal{W} _n(}$$p, \bf{\Sigma}, \bf{\Theta})$, if its pdf is given by \cite[eq. (99)]{James}
\begin{equation}
\begin{split}
f_\mathbf{W}(\mathbf{W})=&\frac{\text{etr} (-\mathbf{\Sigma} ^{-1} \mathbf{W})\vert \mathbf{W} \vert ^{p-n}}{\widetilde{\Gamma} _n(p) \vert \mathbf{\Sigma} \vert ^p} \\ &\times
 \text{etr}(-\mathbf{\Theta}) _0\widetilde{F}_1(p; \mathbf{\Theta} \mathbf{\Sigma} ^{-1} \mathbf{W}),
\end{split}
\end{equation}
where $\widetilde{\Gamma}_n(p)$ is the complex multivariate gamma function \cite[eq. (83)]{James}, and $_0\widetilde{F}_1(\cdot; \cdot)$ is the complex Bessel hypergeometric function of matrix argument \cite {James}. Notice that the first line expression of the eq. (1) corresponds to the pdf of a central complex Wishart matrix \cite[eq. (94)]{James}.

\emph{Definition 2: Complex Gamma-variate Matrix.}\\
The $n\times n$ Hermitian positive-definite matrix $\bf{B}$ is a complex gamma-variate matrix, with scalar parameter $\beta\ (\beta~\geq~n)$ and matrix parameter $\bf{\Omega}$, if $\bf{B}$ follows the complex gamma distribution $\Gamma_n(\beta, \bf{\Omega})$ \cite[p.~254, p. 356]{Mathai}, i.e,
\begin {equation}
f_\mathbf{B}(\mathbf{B})=\frac{\vert \mathbf{B} \vert ^{\beta-n}\vert \mathbf{\Omega} \vert ^{\beta}}{\widetilde{\Gamma} _n(\beta)}\text{etr} (-\mathbf{\Omega} \mathbf{B}).
\end{equation}
Notice that the complex gamma distribution $\Gamma_n(\beta, \bf{\Omega})$ can be seen as the continuous extension of the central Wishart distribution when its scalar parameter takes real positive values.

\emph{Definition 3: Complex Gamma-Wishart Matrix.}\\
Assume to have a $q\times n~(q \geq n)$ matrix, $\mathbf{H}$, defined as
\begin {equation}
\mathbf{H}=\mathbf{\hat{H}}+\mathbf{\bar{H}}
\end {equation}
where $\mathbf{\hat{H}}\sim\mathcal{CN}(0, \mathbf{I}_q\otimes\bf{\Sigma})$ and $\mathbf{\bar{H}}^\mathcal{H}\mathbf{\bar{H}}\sim\Gamma_n(\alpha, \bf{\Omega})$ are statistically independent. Then the Gram matrix $\mathbf{A}=\mathbf{H}^{\cal{H}}\bf{H}$ follows the gamma-Wishart distribution $\Gamma\mathcal{W}_n(\alpha, q, \bf{\Sigma}, \bf{\Omega})$ given by \cite{Alfano, Mckay}
\begin {equation}
\begin{split}
f_\mathbf{A}(\mathbf{A})=&\frac{\text{etr} (-\mathbf{\Sigma}^{-1} \mathbf{A})\vert \mathbf{A} \vert ^{q-n}\vert \mathbf{\Omega} \vert ^{\alpha}}{\widetilde{\Gamma}_n(q)\vert \mathbf{\Sigma} \vert ^{q}\vert \mathbf{\Sigma}^{-1}+\mathbf{\Omega} \vert ^{\alpha}} \\ &\times~_1\widetilde{F}_1(\alpha; q; \mathbf{\Sigma}^{-1}(\mathbf{\Sigma}^{-1}+\mathbf{\Omega})^{-1} \mathbf{\Sigma}^{-1}\mathbf{A}),
\end {split}
\end{equation}
where $_1\widetilde{F}_1(\cdot; \cdot; \cdot)$ is the complex confluent hypergeometric function of matrix argument \cite{James}.

\section{System model definition}

The MIMO Rician shadowed fading model for multi-antenna communications was presented in \cite{Alfano} with the following channel matrix model
\begin{equation}
\mathbf{H}=\mathbf{\hat{H}}+\mathbf{\Xi},
\end{equation}
where $\mathbf{\hat{H}}\sim\mathcal{CN}(0, \mathbf{I}_{p}\otimes\bf{\Sigma})$, and $\mathbf{\Xi}^{\mathcal{H}}\mathbf{\Xi}\sim\Gamma_n(m, \bf{M})$; matrix $\mathbf{\hat{H}}$ represents the scattered components, $\mathbf{\Xi}$ is the LOS which suffers from shadowing and $\mathbf{M}$ takes into account the spatial correlation of the shadowing at the receiver side, since the number of receiver antennas is lower than the number of transmit antennas $p$. Notice that this does not imply any loss of generality, since every Gram matrix will have $n\times n$ elements, with $n=\text{min}(p, n)$, but we take this assumption for the sake of notational simplicity.

The pdf of the Gram channel matrix, $\mathbf{Y}=\mathbf{H}^\mathcal{H}\mathbf{H}$, can be given in terms of the well known gamma-Wishart distribution, such as $\mathbf{Y}\sim\Gamma\mathcal{W}_n(m,p,\mathbf{\Sigma},\mathbf{M})$ \cite[Proposition 1]{Alfano}. Next, the mgf of the MIMO Rician shadowed model is derived.

\emph{Lemma 1}: Let $\mathbf{Y}\sim\Gamma\mathcal{W}_n(m, p, \mathbf{\Sigma}, \mathbf{M})$; then, its mgf is given by
\begin {equation}
\label{mgf}
\begin{split}
\mathcal{M}_\mathbf{Y}(\mathbf{S})\triangleq&~\text{E}\left[\text{etr}(\mathbf{YS})\right]\\
=&\frac{\vert\mathbf{-S+\Sigma}^{-1}\vert^{-p}\vert\mathbf{M}\vert^m}{\vert\mathbf{\Sigma}\vert^{p}\vert\mathbf{\Sigma}^{-1}+\mathbf{M}\vert^{m}} \\ \times&\vert\mathbf{I}_n-\mathbf{\Sigma}^{-1}(\mathbf{\Sigma}^{-1}+\mathbf{M})^{-1}\mathbf{\Sigma}^{-1}(\mathbf{-S+\Sigma}^{-1})^{-1}\vert^{-m}.
\end{split}
\end {equation}
\emph{Proof}: The mgf is calculated from the next integration over the space of Hermitian positive definite matrices
\begin{equation}
\label{integral}
\mathcal{M}_\mathbf{Y}(\mathbf{S})=\int_{\mathbf{Y}^{\mathcal{H}}=\mathbf{Y}>\text{0}}\text{etr}(\mathbf{YS})\cdot \text{\emph{f}}_\mathbf{Y}(\mathbf{Y})(\text{d}\bf{Y})
\end{equation}
where $\text{\emph{f}}_\mathbf{Y}(\mathbf{Y})$ is the pdf of the matrix $\bf{Y}$, which depends on the hypergeometric $_1\widetilde{F}_1(\cdot; \cdot; \cdot)$. Eq.~(\ref{integral}) is carried out with the help of \cite[eq. (6.1.20)]{Mathai}, when expressing the hypergeometric function in series form \cite[eq. (87)]{James}. Thus, a Binomial hypergeometric function $_1\widetilde{F}_0(\cdot; \cdot)$ is obtained and expressed in turn as a determinant \cite[eq. (90)]{James}. This result is new in the literature to the best of our knowledge.

\section{Unification of popular fading models through the MIMO Rician shadowed}
In this section, we first show that the MIMO Rician shadowed model unifies a wide set of SISO models, and then we also show what MIMO models are included therein. The MIMO Rician shadowed model parameters are underlined for the sake of clarity.

\subsection{Unifying models for SISO communication}

In the following lemma, we prove that the $\kappa$-$\mu$ shadowed, originally proposed by \cite{Paris}, can be seen as a particular case of the MIMO Rician shadowed case.

\emph{Lemma 2}: Let $\mathbf{Y}\sim\Gamma\mathcal{W}_n(m, p, \mathbf{\Sigma}, \mathbf{M})$. If $n=1$, then $\mathbf{\Sigma}\rightarrow\sigma_{\Sigma}^2$, $\mathbf{M}\rightarrow\sigma_{M}^2$ and $\mathbf{Y}\rightarrow y$. Let $\gamma=\bar{\gamma}\frac{y}{\bar{y}}$, with $\bar{\gamma}=\mathbb{E}[\gamma]$ and $\bar{y}=\mathbb{E}[y]$, then $\gamma$ follows a \mbox{$\kappa$-$\mu$} shadowed distribution with $\kappa=m\sigma_M^{-2}/(\mu\sigma_\Sigma^{2})$ and $\mu=p$.

\emph{Proof}: If $n=1$, we can identify $\underline{\sigma_{M}}^{2}=m\kappa^{-1}\bar{\gamma}^{-1}(1+\kappa)$, $\underline{\sigma_{\Sigma}}^{-2}=\mu\bar{\gamma}^{-1}(1+\kappa)$, $\underline{p}=\mu$ and $\underline{m}=m$ in the Gamma-Wishart distribution and so we obtain the $\kappa$-$\mu$ shadowed distribution originally presented in \cite[eq.~(4)]{Paris}. Notice also that eq.~(\ref{mgf}) becomes \cite[eq.~(5)]{Paris}.  

Very recently, the $\kappa$-$\mu$ shadowed has been presented as the model which unifies the $\kappa$-$\mu$ and $\eta$-$\mu$ distributions \cite{Laureano}. Therefore, the MIMO Rician shadowed includes the $\kappa$-$\mu$ shadowed, $\kappa$-$\mu$, $\eta$-$\mu$, Nakagami-$m$, Nakagami-$q$, Rician, Rayleigh and one-sided Gaussian.

\subsection{Unifying models for MIMO communication}

The MIMO Rayleigh and MIMO Rician can be deduced from the MIMO Rician shadowed fading model when its parameters are set to specific values and/or taken to limit. Table I summarizes these MIMO fading derivations. Notice that considering a MIMO $\kappa$-$\mu$ or a MIMO $\kappa$-$\mu$ shadowed does not give a different distribution when compared to the MIMO Rician or MIMO Rician shadowed, respectively, since the sum of noncentral Wishart gives another noncentral Wishart with more degree of freedom. While a unification similar to the SISO case under the umbrella of the MIMO Rician shadowed model could be inferred for the Nakagami-$q$ \cite{Laureano}, such connection is not possible when $n>1$. Therefore, exploring this possibility is indeed an interesting and challenging problem.

\begin{table}[!t]
\renewcommand{\arraystretch}{1.7}
\caption{The channels derived from MIMO Rician shadowed model}
\label{table_example}
\centering
\begin{tabular}
{c|c}
\hline
\hline
MIMO Channels (Distributions)  & MIMO Rician Shadowed Parameters\\
\hline
\hline
MIMO Rayleigh &   \b{$\bf{M}$}$^{-1}\rightarrow \mathbf{0}_n$\\ \cline{2-2}
(Central Wishart)&   \b{$m$}~$=$~\b{$p$} \\
\hline
\begin{tabular}{c}MIMO Rician\\(Noncentral Wishart) 
\\ \end{tabular} &  \b{$m$}~$\rightarrow\infty$\\
\hline
\hline
\end{tabular}
\end{table}

Due to space constraints, we only outline the proofs which are required to obtain the results in Table I. On the one hand, the derivation for the MIMO Rayleigh fading model is carried out thanks to the next properties of the hypergeometric functions. When the case $\mathbf{M}^{-1}\rightarrow \mathbf{0}_n$ is considered, we apply
\begin{equation}
\label{property1}
\lim_{c\rightarrow 0}~_p\widetilde{F}_q\left(a_1\ldots a_p; b_1\ldots b_q; c\mathbf{X}\right)=1.
\end{equation}
When $m=p$, we use
\begin{equation}
\label{property2}
_1\widetilde{F}_1\left(a ; a; \mathbf{X}\right)=\text{etr}(\mathbf{X})_1\widetilde{F}_1\left(a-a ; a; -\mathbf{X}\right)=\text{etr}(\mathbf{X}).
\end{equation}
In fact, the eq.~(\ref{property1}) can be handled by simply exploiting the series expression of the hypergeometric function of matrix argument \cite[eq. (87)]{James}, where the first term has the unit value and the rest of the terms depend on the eigenvalues of the matrix argument, which become zero when $\bf{M}$~$\rightarrow \mathbf{0}_n$. The eq.~(\ref{property2}), which is usually referred to as the Kummer relation for scalar confluent hypergeometric functions, is derived by using the integral representation of the hypergeometric function \cite[eq. (6.2.4)]{Mathai}. 

On the other hand, the MIMO Rician case is derived by using the following limits
\begin{equation}
\label{property3}
\lim_{a\rightarrow\infty}~_1\widetilde{F}_1\Big(a; b; \frac{1}{a}\mathbf{X}\Big)=~_0\widetilde{F}_1\Big(b; \mathbf{X}\Big)
\end{equation}
\begin{equation}
\label{property4}
\lim_{m\rightarrow\infty}~\vert\mathbf{I}_n+\frac{1}{m}\mathbf{\Sigma}^{-1}\mathbf{M}^{-1}\vert^{-m}=\text{etr}\left(-\mathbf{\Sigma}^{-1}\mathbf{M}^{-1}\right)
\end{equation}
Eq.~(\ref{property3}) can be proved by expressing the hypergeometric function in series \cite[eq. (87)]{James}. The constant of the zonal polynomial argument can be then extracted from it, so that the complex Pochhammer symbol vanishes when taking the limit. Eq.~(\ref{property4}) can be derived by expressing the determinant as the eigenvalue product $\prod_{i=1}^{n}\left(1+\frac{1}{m}\lambda_i\right)^{-m}$, so that each product component tends to the exponential function.

\section{Maximum eigenvalue distribution}

The study of the maximum eigenvalue distribution depends on the eigenvalues of the matrix $\mathbf{M}$. For a matrix $\mathbf{M}$ with distinct eigenvalues, we find the result in \cite{Alfano}. For a matrix $\mathbf{M}$ with two or more eigenvalues equals, it is not possible to use the result in \cite{Alfano}, since we have an indeterminate form $0/0$. Instead we should derive a new result thanks to the Lemma 2 in \cite{Chiani2010}, which is based on applying multiple times L'H\^opital theorem.

However, we are going to focus here in the simplest case where all the eigenvalues of $\mathbf{M}$ are equals. For that case, we suppose that $\mathbf{M}$ is a diagonal matrix with equals elements. This does not imply any loss of generality since the solution depends on the eigenvalues of $\mathbf{M}$ and not necessarily the shadowing has to be spatially uncorrelated, i.e, $\mathbf{M}$ has not to be a diagonal matrix.

\begin{figure*}[!b] 
\hrulefill
\normalsize
\setcounter{equation}{14}
\begin {equation}
\begin{split}
\label{eqmia}
\{\mathbf{\Upsilon}(x)\}_{i,j}=~&\sigma_{\Sigma}^{2p-2j+2}\Big(1+\sigma_\Sigma^{2}\sigma_M^{2}\Big)^{i-n}\Gamma(\tau-i-j+1)\Bigl[~_2\mathcal{F}_1\Big(\tau-i-j+1, m-i+1; p-i+1; \frac{1}{1+\sigma_\Sigma^{2}\sigma_M^{2}}\Big)\\
&-\text{e}^{-\sigma_\Sigma^{-2}x}\sum_{k=0}^{\tau-i-j}\frac{(\sigma_\Sigma^{-2}x)^k}{k!}\Phi_1\Big(m-i+1,\tau-i-j-k+1,p-i+1,\frac{1}{1+\sigma_\Sigma^{2}\sigma_M^{2}},\frac{\sigma_\Sigma^{-2}x}{1+\sigma_\Sigma^{2}\sigma_M^{2}}\Big)\Bigr]
\end{split}
\end{equation}
\setcounter{equation}{11}
\end{figure*}

\emph{Corollary 1:} The joint distribution of the ordered eigenvalues $\phi_1 < \phi_2 < \ldots < \phi_n$ of $\mathbf{Y}\sim~\Gamma\mathcal{W}_n(m, p, \mathbf{\Sigma}, \mathbf{M})$, when $\mathbf{\Sigma}=\sigma_\Sigma^2\mathbf{I}_n$ and $\mathbf{M}=\sigma_M^2\mathbf{I}_n$  is given by
\begin{equation}
\label{joint}
\begin{split}
f_{\mathbf{\Phi}}(\mathbf{\Phi})=&~\frac{\pi^{n(n-1)}\prod_{i<j}^{n}(\phi_i-\phi_j)^2}{\sigma_\Sigma^{2p n}\widetilde{\Gamma}_n(n)\widetilde{\Gamma}_n(p)\left(1+\sigma_\Sigma^{-2}\sigma_M^{-2}\right)^{nm}}\\
&\times\vert\mathbf{\Phi}\vert^{p-n}\text{etr}(-\sigma_{\Sigma}^{-2}\mathbf{\Phi})_1\widetilde{F}_1\Big(m; p; \frac{\sigma_\Sigma^{-2}\mathbf{\Phi}}{1+\sigma_\Sigma^{2}\sigma_M^{2}}\Big),
\end{split}
\end{equation}
where the confluent hypergeometric function is of one matrix argument and $\mathbf{\Phi}=\text{diag}(\phi_i)$.

\emph{Proof}: Applying \cite[eq. (88)]{James}, the integration over the space of unitary matrices of the pdf of $\mathbf{Y}$ leads to the result. Next, the cdf of the maximum eigenvalue is derived.

\emph{Lemma 3}: Let $\tau=p+n$, the cdf of the maximum eigenvalue of $\mathbf{Y}\sim\Gamma\mathcal{W}_n(m, p, \mathbf{\Sigma},\mathbf{M})$, when $\mathbf{\Sigma}=\sigma_\Sigma^2\mathbf{I}_n$ and $\mathbf{M}=\sigma_M^{2}\mathbf{I}_n$, can be expressed as
\begin{equation}
\label{cdf}
F_{\phi_n} (\phi_n)=C\vert\mathbf{\Upsilon}(\phi_n)\vert,
\end{equation}
where the constant $C$ can be expressed as 
\begin{equation}
C=\frac{\pi^{n(n-1)}\Big[\sigma_\Sigma^2\Big(1+\sigma_\Sigma^{2}\sigma_M^{2}\Big)\Big]^{\frac{n(n-1)}{2}}}{\sigma_\Sigma^{2p n}\widetilde{\Gamma}_n(n)\widetilde{\Gamma}_n(p)\left(1+\sigma_\Sigma^{-2}\sigma_M^{-2}\right)^{nm}}.
\end{equation}
When $m<p$, the entries of the $n\times n$ matrix $\mathbf{\Upsilon}(x)$ are given by the eq.~(\ref{eqmia}) at the bottom of the page, where $_2\mathcal{F}_1$ is the Gauss hypergeometric function of scalar argument \cite[eq. (15.1.1)]{Abramowitz}, $\Phi_1(\cdot,\cdot,\cdot,\cdot)$ is the confluent hypergeometric function of two scalar variables \cite[eq. (9.261.1)]{Gradsteyn}, and $\Gamma(a)$ is the univariate gamma function. When $m>p$, no closed-form expression is obtained, so we give the entries of $\mathbf{\Upsilon}(x)$ in the following integral form
\addtocounter{equation}{1}
\begin {equation}
\label{integral_cdf}
\begin{split}
&\{\mathbf{\Upsilon}(x)\}_{i,j}=\Big[\sigma_\Sigma^{2}\Big(1+\sigma_\Sigma^{2}\sigma_M^{2}\Big)\Big]^{i-n}\Big\{\int_{0}^{x}y^{\tau-i-j}\text{e}^{-\sigma_\Sigma^{-2}y}\\
&\times_1\mathcal{F}_1\Big(m-i+1; p-i+1; \frac{\sigma_\Sigma^{-2}y}{1+\sigma_\Sigma^{2}\sigma_M^{2}}\Big)dy\Big\}.
\end{split}
\end {equation}
Finally, when $m=p$, $\mathbf{Y}$ follows a central Wishart distribution, so that its extreme eigenvalue distributions are given in \cite{Khatri1}.

\emph{Proof}: The cdf of the maximum eigenvalue is derived by integrating the joint eigenvalue distribution in eq. (15) multiple times, such as indicated in \cite[Appendix A]{Alouini}. As shown in \cite[eq. (2.9)]{Richards}, the hypergeometric function of one matrix argument in eq.~(\ref{joint}) can be expressed by a division of determinants, which gives a product of two determinants in eq.~(\ref{joint}). Since the multiple integrals of a product of two determinants can be expressed as a determinant of a single integral \cite{Andreief}, we finally obtain the integral form of eq.~(\ref{integral_cdf}), which can be expressed as a finite sum of confluent hypergeometric functions of two scalar variables when $m < p$.

\emph{Lemma 4}: Let $\tau=p+n$, the pdf of the maximum eigenvalue $\mathbf{Y}\sim\Gamma\mathcal{W}_n(m, p, \mathbf{\Sigma}, \mathbf{M})$, when $\mathbf{\Sigma}=\sigma_\Sigma^2\mathbf{I}_n$ and $\mathbf{M}=\sigma_M^2\mathbf{I}_n$, can be expressed as
\begin {equation}
\begin{split}
f_{\phi_n} (\phi_n)=&C\vert\mathbf{\Upsilon}(\phi_n)\vert\\
&\times\text{tr}\left\{\mathbf{\Upsilon}^{-1}(\phi_n)\mathbf{J}(\phi_n)\right\}\text{U}(\phi_n).
\label{pdf}
\end{split}
\end {equation}
where the $\text{U}(\cdot)$ is the unit step function and the entries of the $n\times n$ matrix $\mathbf{J}(x)$ are given by the derivatives of the entries of $\mathbf{\Upsilon}(x)$ with respect to $x$, i.e., by removing the integral in eq.~(\ref{integral_cdf}).

\emph{Proof}: The proof is straightforward by using the derivative formula of a determinant given by \cite[eq. (9)]{Golberg}. Notice that this can be also applied in the case of distinct eigenvalues presented by \cite{Alfano} to obtain the other case pdf of the maximum eigenvalue. 

\section{Numerical Results}

In order to validate our analytical results, we compare them with Monte-Carlo simulations. Fig. 1 shows different simulated and theoretical curves of the cdf of the maximum eigenvalue when the shadowing has equals eigenvalues. We appreciate a perfect match between simulated and theoretical values. In turn, Fig. 2 allows to validate our theoretical expression for the pdf of the maximum eigenvalue.

Interestingly, the MIMO Rician shadowed model is more flexible than other existing MIMO models. Fig. 3 shows the evolution of the pdf of the maximum eigenvalue as the parameter $m$ grows, which prove that, when $m\rightarrow\infty$, the MIMO Rician shadowed maximum eigenvalue distribution tends to the MIMO Rician one, which was presented in \cite{Alouini}. We can also observe that we have a wide set of possible pdf of the maximum eigenvalue, which lay between the MIMO Rician case and the Rayleigh case for $m\geq p$. Also, when $m<p$, we obtain pdfs which correspond to channels that present a fading more severe than the Rayleigh case, often called hyper-Rayleigh channels \cite{Frolik}.

\begin{figure}[!t]

\centering
\pgfplotsset{every axis/.append style={
xmin={0},
xmax={180},
ymin={0},
ymax={1},
}}
\begin{tikzpicture}[scale=0.8]
\begin{axis}[
grid=both,
width=11cm,
height=5.8cm, 
xlabel=$\phi_n$,
ylabel=$F_{\phi_n}(\phi_n)$,
legend cell align=left,
legend entries={$n=2$\text{,} $m=2$\text{,} $\sigma_M^{-2}=8$\text{,} $\sigma_\Sigma^2=1$, $n=3$\text{,} $m=3$\text{,} $\sigma_M^{-2}=8$\text{,} $\sigma_\Sigma^2=1$, $n=2$\text{,} $m=2$\text{,} $\sigma_M^{-2}=40$\text{,} $\sigma_\Sigma^2=1$, $n=3$\text{,} $m=3$\text{,} $\sigma_M^{-2}=8$\text{,} $\sigma_\Sigma^2=4$, Monte-Carlo},
legend style={font=\footnotesize, legend pos=south east},
legend columns=1
]
\addplot[color=red, thick, densely dotted] table[x=x,y=y] {cdfTeorican2m2mu1p4tau0p5s1maspt.dat};
\addplot[color=red,thick, densely dashed] table[x=x,y=y] {cdfTeorican3m3mu1p4tau0p5s1maspt.dat};
\addplot[color=red,thick, dash pattern=on 4pt off 4pt on 1pt off 4 pt] table[x=x,y=y] {cdfTeorican2m2mu1p4tau0p1s1maspt.dat};
\addplot[color=red, thick] table[x=x,y=y] {cdfTeorican3m3mu1p4tau0p5s2.dat};

\addplot[color=blue,only marks, mark=o] table[x=x,y=y] {cdfn2m2p4tau0p5s1menospt.dat};
\addplot[color=blue,only marks, mark=o] table[x=x,y=y] {cdfn3m3p4tau0p5s1menospt.dat};
\addplot[color=blue,only marks, mark=o] table[x=x,y=y] {cdfn2m2p4tau0p1s1menospt.dat};
\addplot[color=blue, only marks, mark=o] table[x=x,y=y] {cdfn3m3p4tau0p5s2menospt.dat};

\end{axis}
\end{tikzpicture}
\caption{Comparison of the analytical and simulated cdf of the maximum eigenvalue of the MIMO Rician shadowed model for different matrix dimensions and various parameters. For all the cases $p=4$.}
\end{figure}
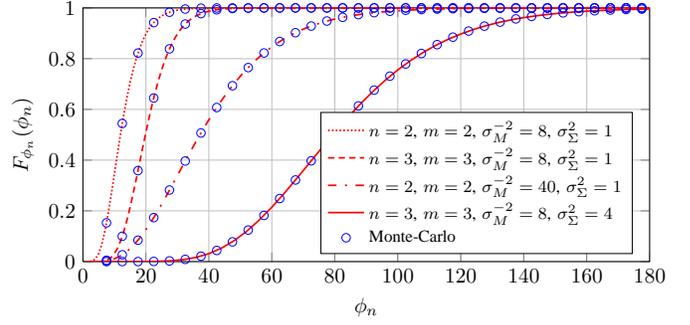

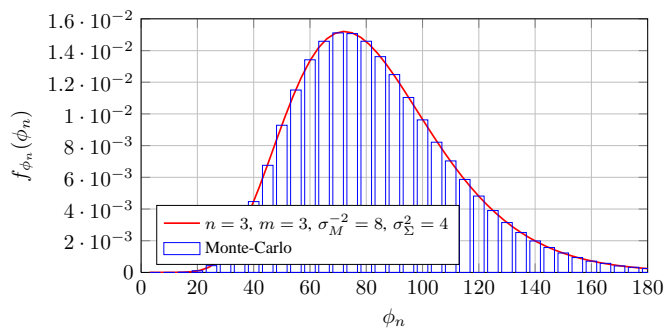
\begin{figure}[!t]
\pgfplotsset{
legend image code/.code={
\draw[#1] (0cm,-0.1cm) rectangle (0.6cm,0.1cm);
}
}
\centering
\pgfplotsset{every axis/.append style={
xmin={0},
xmax={180},
ymin={0},
ymax={0.016},
ytick={0,0.002,...,0.016}
}}
\begin{tikzpicture}[scale=0.8]
\begin{axis}[
grid=both,
width=10cm,
height=5.8cm, 
xlabel=$\phi_n$,
ylabel=$f_{\phi_n}(\phi_n)$,
legend cell align=left,
legend entries={$n=3$\text{,} $m=3$\text{,} $\sigma_M^{-2}=8$\text{,} $\sigma_\Sigma^2=4$, Monte-Carlo},
legend style={font=\footnotesize, legend pos=south west},
legend columns=1,
bar width=5pt
]
\addplot[color=red, line legend, thick] table[x=x,y=y] {pdfTeorican3m3mu1p4tau0p5s2.dat};
\addplot[ybar, mark=none, draw=blue] table[x=x,y=y] {pdfn3m3p4tau0p5s2.dat};

\end{axis}
\end{tikzpicture}

\caption{Analytical and simulated pdf of the MIMO Rician shadowed maximum eigenvalue for the last case of Fig. 1.}
\label{PDF}
\end{figure}

\begin{figure}[!t]

\centering
\pgfplotsset{every axis/.append style={
xmin={0},
xmax={40},
ymin={0},
ymax={0.09},
ytick={0,0.01,...,0.09}
}}
\begin{tikzpicture}[scale=0.8]
\begin{axis}[
grid=both,
width=10cm,
height=5.8cm, 
xlabel=$\phi_n$,
ylabel=$f_{\phi_n}(\phi_n)$,
legend cell align=left,
legend entries={$m\rightarrow\infty$ $[$30$]$, $m=100$, $m=50$, $m=10$, $m=4$, $m=2$},
legend style={font=\footnotesize, legend pos=north west},
legend columns=1
]
\addplot[color=red, thick] table[x=x,y=y] {pdfTeoricanAlouni.dat};
\addplot[color=blue,thick, densely dotted] table[x=x,y=y] {pdfTeorican2m100mu1p4kappa10s1.dat};
\addplot[color=blue,thick, dash pattern=on 6pt off 6pt] table[x=x,y=y] {pdfTeorican2m50mu1p4kappa10s1.dat};
\addplot[color=blue,thick, dash pattern=on 4pt off 4pt on 1pt off 4pt] table[x=x,y=y] {pdfTeorican2m10mu1p4kappa10s1.dat};
\addplot[color=blue,thick] table[x=x,y=y] {pdfTeorican2m4mu1p4kappa10s1.dat};
\addplot[color=blue,thick,dash pattern=on 1pt off 7pt] table[x=x,y=y] {pdfTeorican2m2mu1p4kappa10s1.dat};

\draw [stealth-,thick] (axis cs:18,0.081) -- (axis cs:24,0.085) node[right]{\footnotesize MIMO Rician};

\draw [stealth-,thick] (axis cs:18,0.053) -- (axis cs:24,0.062) node[right]{\footnotesize MIMO Rayleigh};

\end{axis}
\end{tikzpicture}

\caption{Evolution of the pdf of the MIMO Rician shadowed maximum eigenvalue when $m$ grows. The other parameters are fixed to $n=2$, $p=4$ and $\sigma_\Sigma^2=1$, with $m\cdot\sigma_M^{-2}=40$.}
\label{conv}
\end{figure}
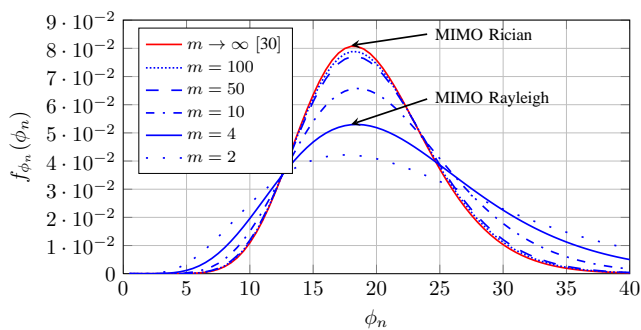

\section{Conclusions}
We have shown that the MIMO Rician shadowed model exhibits a powerful unification, since it includes the MIMO Rayleigh and MIMO Rician, together with the SISO $\kappa$-$\mu$ shadowed, $\kappa$-$\mu$, $\eta$-$\mu$, Nakagami-$m$, Nakagami-$q$, Rician, Rayleigh and one-sided Gaussian. Therefore, it gives more flexibility to model any channel affected by different propagation conditions with a tractable statistic characterization than existing alternatives. Furthermore, the mgf of the Gram channel matrix and the cdf and pdf of its maximum eigenvalue for a shadowing with equal eigenvalues have been obtained in closed-form. 


\section*{Acknowledgment}

This work has been funded by Consejer\'ia de Econom\'ia,
Innovaci\'on, Ciencia y Empleo - Junta de Andaluc\'ia, Universidad de M\'alaga - CEI Andaluc\'ia TECH, the Spanish Government and the European Regional Development Fund (projects P2011-TIC-7109, P2011-TIC-8238, TEC2011-25473 and COFUND2013-40259).



%

\bibliographystyle{IEEEtran}
\bibliography{bibfile}

\begin{thebibliography}{10}
\providecommand{\url}[1]{#1}
\csname url@samestyle\endcsname
\providecommand{\newblock}{\relax}
\providecommand{\bibinfo}[2]{#2}
\providecommand{\BIBentrySTDinterwordspacing}{\spaceskip=0pt\relax}
\providecommand{\BIBentryALTinterwordstretchfactor}{4}
\providecommand{\BIBentryALTinterwordspacing}{\spaceskip=\fontdimen2\font plus
\BIBentryALTinterwordstretchfactor\fontdimen3\font minus
  \fontdimen4\font\relax}
\providecommand{\BIBforeignlanguage}[2]{{%
\expandafter\ifx\csname l@#1\endcsname\relax
\typeout{** WARNING: IEEEtran.bst: No hyphenation pattern has been}%
\typeout{** loaded for the language `#1'. Using the pattern for}%
\typeout{** the default language instead.}%
\else
\language=\csname l@#1\endcsname
\fi
#2}}
\providecommand{\BIBdecl}{\relax}
\BIBdecl

\bibitem{Rician}
S.~O. Rice, ``Statistical properties of a sine wave plus random noise,''
  \emph{Bell Syst. Tech. J.}, vol.~27, no.~1, pp. 109--157, 1948.

\bibitem{Loo}
C.~Loo, ``A statistical model for a land mobile satellite link,'' \emph{IEEE
  Trans. Veh. Technol.}, vol.~34, no.~3, pp. 122--127, Aug 1985.

\bibitem{Abdi}
A.~Abdi, W.~Lau, M.-S. Alouini, and M.~Kaveh, ``A new simple model for land
  mobile satellite channels: first- and second-order statistics,'' \emph{IEEE
  Trans. Wireless Commun.}, vol.~2, no.~3, pp. 519--528, May 2003.

\bibitem{Paris}
J.~Paris, ``Statistical characterization of $\kappa $ - $\mu$ shadowed
  fading,'' \emph{IEEE Trans. Veh. Technol.}, vol.~63, no.~2, pp. 518--526, Feb
  2014.

\bibitem{CottonShadowed}
S.~L. Cotton, ``Human body shadowing in cellular device-to-device
  communications: Channel modeling using the shadowed fading model,''
  \emph{IEEE J. Sel. Areas Commun.}, vol.~33, no.~1, pp. 111--119, 2015.

\bibitem{ParisConf}
A.~S{\'a}nchez, E.~Robles, F.~Rodrigo, F.~Ruiz-Vega, U.~Fern{\'a}ndez-Plazaola,
  and J.~Paris, ``Measurement and modelling of fading in ultrasonic underwater
  channels,'' in \emph{Proc. Underwater Acoustic Conf.}, 2014.

\bibitem{Cotton}
S.~L. Cotton, ``Shadowed fading in body-to-body communications channels in an
  outdoor environment at 2.45 ghz,'' in \emph{2014 IEEE-APS Topical Conf.
  Antennas Propag. Wireless Commun. (APWC)}, 2014.

\bibitem{WintersCo}
J.~Winters, ``Optimum combining in digital mobile radio with cochannel
  interference,'' \emph{IEEE J. Sel. Areas Commun.}, vol.~2, no.~4, pp.
  528--539, July 1984.

\bibitem{WintersCapacity}
------, ``On the capacity of radio communication systems with diversity in a
  rayleigh fading environment,'' \emph{Selected Areas in Communications, IEEE
  Journal on}, vol.~5, no.~5, pp. 871--878, Jun 1987.

\bibitem{Telatar}
I.~E. Telatar \emph{et~al.}, ``Capacity of multi-antenna gaussian channels,''
  \emph{European transactions on telecommunications}, vol.~10, no.~6, pp.
  585--595, 1999.

\bibitem{ChianiRayleigh}
M.~Chiani, M.~Win, and A.~Zanella, ``On the capacity of spatially correlated
  mimo rayleigh-fading channels,'' \emph{IEEE Trans. Inf. Theory}, vol.~49,
  no.~10, pp. 2363--2371, Oct 2003.

\bibitem{Khalighi}
M.-A. Khalighi, J.~Brossier, G.~Jourdain, and K.~Raoof, ``On capacity of rician
  mimo channels,'' in \emph{Personal, Indoor and Mobile Radio Communications,
  2001 12th IEEE International Symposium on}, vol.~1, Sep 2001, pp.
  A--150--A--154 vol.1.

\bibitem{MckayBounds}
M.~McKay and I.~Collings, ``General capacity bounds for spatially correlated
  rician mimo channels,'' \emph{Information Theory, IEEE Transactions on},
  vol.~51, no.~9, pp. 3121--3145, Sept 2005.

\bibitem{AlouiniCapacityRician}
M.~Kang and M.-S. Alouini, ``Capacity of mimo rician channels,'' \emph{IEEE
  Trans. Wireless Commun.}, vol.~5, no.~1, pp. 112--122, Jan 2006.

\bibitem{MIMO2x2}
G.~Fraidenraich, O.~L{\'e}v{\^e}que, and J.~Cioffi, ``On the mimo channel
  capacity for the dual and asymptotic cases over hoyt channels,'' \emph{IEEE
  Commun. Lett.}, 2007.

\bibitem{Kumar}
S.~Kumar and A.~Pandey, ``Random matrix model for nakagami--hoyt fading,''
  \emph{IEEE Trans. Inf. Theory}, vol.~56, no.~5, pp. 2360--2372, 2010.

\bibitem{Alfano}
G.~Alfano, A.~De~Maio, and A.~M. Tulino, ``A theoretical framework for lms mimo
  communication systems performance analysis,'' \emph{IEE Trans. Inf. Theory},
  vol.~56, no.~11, pp. 5614--5630, 2010.

\bibitem{Mckay}
S.~Jin, M.~McKay, X.~Gao, and I.~Collings, ``Mimo multichannel beamforming: Ser
  and outage using new eigenvalue distributions of complex noncentral wishart
  matrices,'' \emph{IEEE Trans. Commun.}, vol.~56, no.~3, pp. 424--434, Mar.
  2008.

\bibitem{Yacoub}
M.~Yacoub, ``The $\kappa$-$\mu$ distribution and the $\eta$-$\mu$
  distribution,'' \emph{IEEE Antennas Propag. Mag.}, vol.~49, no.~1, pp.
  68--81, Feb 2007.

\bibitem{Tse2000}
C.-H. Tse, K.-W. Yip, and T.-S. Ng, ``Performance tradeoffs between maximum
  ratio transmission and switched-transmit diversity,'' in \emph{Personal,
  Indoor and Mobile Radio Communications, 2000. PIMRC 2000. The 11th IEEE
  International Symposium on}, vol.~2, 2000, pp. 1485--1489 vol.2.

\bibitem{Dighe2003}
P.~Dighe, R.~Mallik, and S.~Jamuar, ``Analysis of transmit-receive diversity in
  rayleigh fading,'' \emph{IEEE Trans. Commun.}, vol.~51, no.~4, pp. 694--703,
  April 2003.

\bibitem{James}
A.~T. James, ``Distributions of matrix variates and latent roots derived from
  normal samples,'' \emph{Ann. Math. Statist.}, vol.~35, no.~2, pp. 475--501,
  Jun. 1964.

\bibitem{Mathai}
A.~M. Mathai, \emph{Jacobians of matrix transformations and functions of matrix
  argument}.\hskip 1em plus 0.5em minus 0.4em\relax World Scientific, 1997.

\bibitem{Laureano}
L.~Moreno-Pozas, F.~J. Lopez-Martinez, J.~F. Paris, and E.~Martos-Naya, ``The
  $\kappa$-$\mu$ shadowed fading model: Unifying the $\kappa$-$\mu$ and
  $\eta$-$\mu$ distributions,'' \emph{arXiv preprint arXiv:1504.05764}, 2015.

\bibitem{Chiani2010}
M.~Chiani, M.~Win, and H.~Shin, ``Mimo networks: The effects of interference,''
  \emph{IEEE Trans. Inf. Theory}, vol.~56, no.~1, pp. 336--349, Jan 2010.

\bibitem{Abramowitz}
M.~Abramowitz, I.~A. Stegun \emph{et~al.}, \emph{Handbook of mathematical
  functions}.\hskip 1em plus 0.5em minus 0.4em\relax Dover New York, 1972,
  vol.~1, no.~5.

\bibitem{Gradsteyn}
I.~S. Gradshteyn and I.~Ryzhik, ``Table of integrals, series, and products,''
  \emph{USA: Academic}, 2007.

\bibitem{Khatri1}
C.~Khatri, ``Distribution of the largest or the smallest characteristic root
  under null hypothesis concerning complex multivariate normal populations,''
  \emph{The Annals of Mathematical Statistics}, pp. 1807--1810, 1964.

\bibitem{Alouini}
M.~Kang and M.-S. Alouini, ``Largest eigenvalue of complex wishart matrices and
  performance analysis of mimo mrc systems,'' \emph{IEEE J. Sel. Areas
  Commun.}, vol.~21, no.~3, pp. 418--426, 2003.

\bibitem{Richards}
D.~S.~P. Richards, ``Totally positive kernels, p{\'o}lya frequency functions,
  and generalized hypergeometric series,'' \emph{Linear Algebra and its
  Applications}, vol. 137, pp. 467--478, 1990.

\bibitem{Andreief}
C.~Andreief, ``Note sur une relation les integrales definies des produits des
  fonctions". mem. de la soc. sci. bordeaux, 2,'' 1883.

\bibitem{Golberg}
M.~Golberg, ``The derivative of a determinant,'' \emph{American Mathematical
  Monthly}, vol.~79, pp. 1124--1126, Dec. 1972.

\bibitem{Frolik}
J.~Frolik, ``On appropriate models for characterizing hyper-rayleigh fading,''
  \emph{IEEE Trans. Wireless Commun.}, vol.~7, no.~12, pp. 5202--5207, December
  2008.

\end{thebibliography}

\end{document}